\documentclass[a4paper]{aa}
\bibpunct{(}{)}{;}{a}{}{,} 
\usepackage{graphicx}

\usepackage{txfonts}
\usepackage{hyperref}
\usepackage[normalem]{ulem}
\usepackage{subcaption}

\makeatletter
\let\linenumbers\relax
\makeatother

\begin{document}

\title{A magnetar powers the luminous supernova 2023pel, associated with a long gamma-ray burst}
\titlerunning{A magnetar powers the luminous supernova 2023pel associated with a long GRB}

   \author{L.M. Roman Aguilar\inst{1,2} \and M.M. Saez\inst{3,4} \and K. Ertini\inst{1,2} \and M.C. Bersten\inst{1,2,5}
   }
          
    \institute{Facultad de Ciencias Astronómicas y Geofísicas Universidad Nacional de La Plata, Paseo del Bosque S/N B1900FWA, La Plata, Argentina\label{inst1} \and Instituto de Astrofísica de La Plata, CONICET, Argentina\label{inst2} \and RIKEN Center for Interdisciplinary Theoretical and Mathematical Sciences (iTHEMS), Wako, Saitama 351-0198, Japan\label{inst3} \and Department of Physics, University of California, Berkeley, CA 94720\label{inst4} \and Kavli Institute for the Physics and Mathematics of the Universe (WPI),The University of Tokyo. Institutes for Advanced Study, The University of Tokyo, Kashiwa, Chiba 277-8583, Japan\label{inst5}}

   \date{Received /
   Accepted}

 \abstract
 {}{We explore supernova (SN)~2023pel, the most recent event associated with gamma-ray bursts (GRBs), specifically GRB~230812B. SN~2023pel shows high luminosity ($\sim 1.5 \times 10^{43}$ erg s$^{-1}$ at the peak) and low expansion velocities ($\rm{v}\sim 16000\,\rm{km\, s^{-1}}$ at the peak) as compared to other GRB-SNe. These properties seem difficult to reconcile with a single nickel power source. We search for models that can explain the properties of this event.}
 {We calculated a grid of hydrodynamic models based on pre-SN structures derived from evolutionary calculations. We compared our models with observations of SN~2023pel and selected our preferred model using statistical analysis, considering both light curves and expansion velocities. This allowed us to derive a set of physical properties for SN~2023pel.} 
{Our models suggest that the most probable scenario involves a millisecond magnetar as the primary power source, supplemented by energy from radioactive decay.
Specifically, our preferred model includes a spin period of $\rm{P}= 3.2 \ ms$, a magnetic field of $\rm{B}= 28 \times 10^{14} \ G$, explosion energy of 2.3 foe, a nickel mass of $\rm{M}_{\rm{Ni}}= 0.24 \ M_{\odot}$, and an ejected mass of 3.4 $\rm{M}_{\odot}$. Alternatively, we found that a purely nickel-powered model also provides a good match with the observations, though M$_{\rm{Ni}}\geq 0.8 \, \rm{M}_{\odot}$ are always required.
The combination of such high values of $\rm{M}_{\rm{Ni}}$ and low M$_{\rm{ej}}$ are difficult to reconcile, indicating that this scenario is less probable. 
We have also identified a specific region within the peak luminosity-velocity plane where an additional energy source beyond nickel may be necessary to power supernovae with characteristics similar to SN~2023pel.} 
{Our study indicates that an additional energy source beyond radioactive decay is essential to explain the high brightness and relatively low expansion velocities of SN 2023pel. A magnetar-powered model aligns well with these characteristics, similar to the models proposed for the very luminous GRB-SN 2011kl.} 


   \maketitle
%

\section{Introduction}
Supernovae (SNe) represent some of the most energetic and fascinating events in the Universe. These powerful explosions are grouped into different types, often based on the presence or absence of certain spectral lines. Type Ic SNe, characterized by the absence of hydrogen and helium in their spectra, are particularly intriguing, among other
reasons, because a subgroup of these has been linked to long-duration gamma-ray bursts (GRBs). 
The SNe associated with GRBs (GRB-SNe) exhibit more energetic explosions than regular SNe Ic, as indicated by their broad spectral lines (BL) \citep{Woosley:2006,Soderberg:2006, Kelly:2008,Drout:2011}. 
However, there are also some SNe Ic-BL that lack GRB detection (see e.g. SN~2002ap \citep{Foley:2003}, SN~2003jd \citep{Valenti:2008}, SN~2009bb \citep{Pignata:2011}, PTF10qts \citep{Taddia:2019}, iPTF14dby \citep{Taddia:2019}, iPTF15dqg \citep{Taddia:2019}, SN~2020bvc \citep{Ho:2020}). Most SNe Ic-BL would be expected not to show GRBs due to viewing angle effects, even if they all have jets. It is worth noting that deep non-thermal observations (radio and X-ray) can place stronger constraints on the presence of jets in these objects due to ruling out off-axis afterglow emission (e.g. \cite{Corsi:2023}).

The mechanism behind the formation of these objects is still unknown, as is the reason why only a subset of SNe Ic-BL has been associated with GRB detection. Several formation scenarios have been proposed, with the collapsar model \citep{Macfadyen:1999, Macfadyen:2001} and the magnetar model \citep{Usov:1992, Metzger:2007, Bucciantini:2008, Metzger:2011} being among the most extensively studied. 

Valuable insights into their origins can be provided by the study of the SNe rates.
Particularly, the GRB-SNe are very uncommon, representing no more than $\sim 3\%$ among stripped-envelope SNe (SE SNe)  \citep{Shivvers:2017,Marongiu:2019}.
Their rarity is probably an intrinsic feature, although selection effects may also be playing an important role. In any case, given the scarcity of GRB-SNe detected, each new object identified within this class can make a significant contribution to further understanding their origins.

SN~2023pel is the most recent GRB-SN detected. It was discovered on August 13, 2023, at 03:34:56.997 UT by the Zwicky Transient Facility, and it was classified as a Type Ic-BL \citep{Agui:2023}. Approximately 8.5 hours earlier, its associated GRB 230812B was detected by the Fermi Gamma-Ray Burst Monitor (GBM; \citealp[]{Meegan:2009}). The time duration of the GRB ($\rm{T}_{90}$) was about $3\,\rm{s}$ \citep{Roberts:2023}, positioning itself near the limit of the division between short-GRBs and long-GRBs ($t \sim 2\,\rm{s}$). At the redshift measured by \cite{deUgartePostigo:2023} (z$\,=\,$0.36), the isotropic energy release and the isotropic equivalent average luminosity of the burst were computed to be $\rm{E_{\gamma,iso}}\sim 1.15 \times 10^{53}\,\rm{erg}$ and $\rm{L_{\gamma,iso}}\sim 8.8\,\times\,10^{52}\,\rm{erg}$ \citep[(hereafter S24]{Gokul:2024}. Although this isotropic energy is consistent with the energy range observed for long-GRBs, it is among the most energetic and luminous objects in the LGRBs population (S24,\cite{Hussenot-Desenonges:2024}). Additionally, the optical counterpart was also reported by the Global MASTER-Net robotic telescopes network at the same location \citep{Lipunov:2023}.

Recent studies have provided insights into GRB~230812B and the associated SN~2023pel. \citealp[(hereafter HD24)]{Hussenot-Desenonges:2024} conducted a multi-band analysis, gathering over 80 observations in X-ray, ultraviolet, optical, infrared, and sub-millimeter bands through the GRANDMA (Global Rapid Advanced Network for Multi-messenger Addicts) collaboration. They spectroscopically confirmed the presence of an associated supernova, SN 2023pel, and derived a photospheric velocity of $\rm{v}_{\rm{ph}} \sim 17000 \pm 3000\, \rm{km \, s^{-1}}$ near maximum luminosity. Additionally, they compared the luminosity of SN~2023pel with the well-known GRB-SN~1998bw and found that they have similar luminosities, but SN 2023pel exhibits a faster light curve (LC) evolution. However, their comparison was based on pseudo-bolometric LC, using only grizJ bands for SN 2023pel.

S24 also studied the optical counterpart of GRB 230812B and confirmed the presence of a SN Ic-BL using spectroscopic observations. Subsequently, they analyzed the bolometric LC of SN 2023pel, employing an analytic radioactive heating model \citep{Arnett:1982}, deriving a nickel mass of M$_{\rm{Ni}} = 0.38$ M$_{\odot}$, an ejected mass of approximately $\rm{M}_{\rm{ej}} \sim 1$ M$_\odot$ and an explosion energy of $1.3 \times 10^{51}$ erg. The authors also explored the possibility of explaining this object assuming solely a millisecond magnetar to power both the GRB and the SN itself, using an analytical model provided by \cite{Cano:2016iuo}. However, they disfavor this scenario due to the inability to derive a satisfactory model for the optical LC and the X-ray observations simultaneously (see further discussion in Sect. \ref{disc}). Nonetheless, the authors suggest that an alternative solution could involve considering a magnetar model supplemented by an additional energy source--a scenario that is further explored in this work.

Previous studies suggest that SN 2023pel is a GRB-SNe with normal brightness. However, our bolometric calculations indicate that this SN stands out in terms of luminosity within the GRB-SNe population (see details in Sect. \ref{disc}). 
Additionally, the parameters found in S24 for SN 2023pel -- such as relatively low nickel mass, ejected mass, and explosion energy compared to other GRB-SNe -- caught our attention. Motivated by these findings, and the absence of alternative published models, we propose to explore our own hydrodynamic models to analyze the SN 2023pel. We seek to derive an independent set of physical parameters for this SN by considering the possibility of combined energy sources, like radioactive decay and magnetar spin-down. The association between highly luminous SNe such as Superluminous SNe (SLSNe), SNe Ic-BL, or GRB-SNe with magnetars has been explored in the literature \citep{Dessart:2012,Kashiyama:2016,Margalit:2018,Suzuki:2021,Afsariardchi:2020glh,Shankar:2021obg,Song:2023qdv}. In particular, magnetar models have been proposed for SN 2011kl \citep{Greiner:2015, Bersten:2016,Kann:2016ndk}, the most luminous GRB-SN detected to date.

The paper is organized as follows:
In Sect. \ref{sec:Lbol}, we present our bolometric LC calculation for SN 2023pel using the photometric data from S24, after applying respective K-corrections. Sect. \ref{hidro} introduces the hydrodynamic code and the pre-SN models utilized. In Sect. \ref{sec:results} our hydrodynamical modeling is presented, which considers both the LC and expansion velocity evolution to constrain the model parameters. This approach allows us to reduce the existent degeneracy between the model free parameters and leads to more reliable physical parameters estimates \citep{Mazzali:2006,Martinez:2020}.  In Sec. \ref{disc} we discuss our results and present a comparison between the SN 2023pel and other energetic events.
Finally, in Sect. \ref{sec:conclus} our main conclusions are summarized.

\section{Bolometric light curve and expansion velocities}\label{sec:Lbol}

Given the lack of a K-corrected bolometric LC of SN~2023pel in the literature, we undertook our own bolometric calculations. 
We used photometry from S24 observed in the $r$- and $i$-bands, corrected by Galactic extinction and with the GRB afterglow subtracted. These data comprehend 10 epochs for the $r$-band and 9 epochs for the $i$-band.
The distance was calculated based on the reported redshift ($z=0.36$), using a flat $\Lambda$CDM cosmology with \mbox{$\Omega_m = 0.286$} and \mbox{H$_0= 69.6 \, \text{km} \, \text{s}^{-1} \, \text{Mpc}^{-1}$} (S24). 

Due to the high redshift value of SN 2023pel, we computed K-corrections. As the available photometry spans from 4 days to 40 days post-explosion, it was crucial to have spectra within that temporal coverage to accurately determine the K-corrections. HD24 and S24 presented spectra at 1.1, 12, 15, and 15.5 days after the explosion. However, the bolometric LC was constructed using photometric data extending to approximately 37 days after the explosion. Therefore, the gaps between epochs were too large to rely on extrapolations. To address this, we used spectra from SN~2002ap, which exhibits similar behavior to SN~2023pel at those epochs. We gathered 25 spectra of SN~2002ap from the WISeREP\footnote{\url{https://www.wiserep.org/}} archive \citep{Yaron:2012}, from the works of \citet{Gal-Yam:2002}, \citet{Foley:2003}, and \citet{Modjaz:2014}. First, we corrected the spectra of SN~2002ap by redshift and extinction and took them to the redshift and extinction values of SN~2023pel. Then we used the SuperNova Algorithm for K-correction Evaluation \citep[SNAKE;][]{Inserra:2018} to calculate the K-corrections. The results are shown in Table \ref{tab:kcorr} to facilitate their use in other possible objects with similar characteristics. We subsequently interpolated the K-corrections to apply them to the photometry in the available epochs. After applying the cross-filter K-correction to the photometry, the observed $r$-band is converted into the rest-frame $g$-band, and the observed $i$-band is converted into the rest-frame $r$-band. 

\begin{table}
\centering
	\caption{K-corrections estimated for r-band and i-band for SN~2023pel. }
	\label{tab:kcorr}
	\begin{tabular}{lcc} 
		\hline
		Phase (d) & K correction $r$-band & K correction $i$-band \\
		\hline
            4.6  & -0.165 $\pm$ 0.018 & -0.295 $\pm$ 0.007 \\
            12.6 & -0.125 $\pm$ 0.010 & -0.266 $\pm$ 0.006\\
            17.6 & -0.079 $\pm$ 0.012 & -0.162 $\pm$ 0.006\\
            19.6 & -0.129 $\pm$ 0.011 & -0.195 $\pm$ 0.006\\
            26.6 & -0.125 $\pm$ 0.012 & -0.054 $\pm$ 0.006\\
            28.6 & -0.088 $\pm$ 0.012 &  0.009 $\pm$ 0.007\\
            37.6 & -0.163 $\pm$ 0.015 & -0.030 $\pm$ 0.006\\
            40.6 & -0.179 $\pm$ 0.010 & -0.086 $\pm$ 0.006\\
    	\hline
	\end{tabular}
    \tablefoot{The phases are referred to the time of the explosion considered as 60168.79 MJD.}
\end{table}

Once we had the K-corrected photometry in the $g$- and $r$-bands, we determined the bolometric luminosity by applying the calibrations given by \cite{Lyman:2014}, for stripped-envelope SNe. These calibrations establish a tight correlation between bolometric corrections and optical colors, through a parametric approach based on BVRI and $griz$ photometry. The resulting bolometric LC is shown with red dots in the figures of the following sections and is used in our modeling. The times are referenced to the epoch of the maximum bolometric luminosity. For SN 2023pel, this epoch corresponds to \mbox{60181.86 MDJ}, determined by applying a polynomial fit to the observed data (S24, HD24)\footnote{In Fig. \ref{fig:model_2023pel}, we have also considered 55916.85 MJD and 50943.41 MJD as the epoch of the maximum bolometric luminosity for SN 2011kl and SN 1998bw, respectively \citep{Greiner:2015, Clocchiatti:2011}.}. Based on our calculations, the LC of SN 2023pel appears to be more luminous than other GRB SNe (refer to Sect. \ref{sec:comparison} for further comparison), reaching a peak luminosity of L$_{\rm{peak}}\sim 1.5 \times 10^{43} \, \rm{erg \, s^{-1}}$ at \mbox{t$_{\rm{peak}} \sim 13$ days} since explosion. In this paper, we consider the time of the explosion coincident with the GRB detection time, that is, $\rm{t}_{\rm exp}=60168.79 \, \rm{MJD}$.

The hydrodynamical modeling can be additionally constrained by using estimations of the velocity at the photosphere, as it evolves with time. The Fe II line has been proposed as a dependable indicator of the photospheric velocity \citep{Dessart:2005, Schulze:2014}. For this reason, we have gathered 4 velocity measurements of the Fe II line ($\lambda$ 5169 \AA) at different epochs, from the existing literature. From HD24, we extracted two velocity measurements obtained from the spectrum at 12.12 and 15.12 days since the GRB detection time. We also included the velocities calculated by S24. These data are shown in red dots at the bottom panels in Figs. \ref{fig:Ni_model}, \ref{fig:mag_model_1}, \ref{fig:mag_model_2} and \ref{fig:model_2023pel}, where the times are referenced to the epoch of the maximum bolometric luminosity of SN 2023pel. The earliest data point exhibits a significant dispersion as it was obtained through an extrapolation process (see S24 for more details). From Fig. \ref{fig:model_2023pel}, we note that the velocities of SN~2023pel are much lower than those observed for SN~1998bw. These velocities fall within a range between those estimated for GRB-SNe and the SLSNe, resulting very similar to those estimated for the luminous SN~2011kl.

\section{Hydrodynamic modeling}\label{hidro}

To study the nature of SN 2023pel, we compare the LC and velocity data presented in Sect.~\ref{sec:Lbol} with a set of hydrodynamic models applied to initial structures derived from stellar evolutionary calculations. To simulate the explosion of the SN, we use a 1D Lagrangian hydrodynamic code, as described in \cite{Bersten:2011}. 
The code numerically solves the hydrodynamical equations, assuming spherical symmetry for a self-gravitating gas and using the Rosseland mean opacity from OPAL tables \citep[and references therein]{Iglesias:1996} and the molecular tables from \cite{Alexander:1994} for lower temperatures (T $\lesssim$ 8000 K). For radiation transport, the diffusion approximation is employed for optical photons, while grey transport is used for gamma photons generated by the radioactive decay of $^{56}$Ni $\rightarrow$ $^{56}$Co $\rightarrow$  $^{56}$Fe.
Our code allows assuming any $^{56}$Ni\footnote{Hereafter, for simplicity, we will refer to the radioactive nickel $^{56}$Ni as Ni.} distribution and calculates the gamma-ray deposition in each point of the SN ejecta, assuming a constant opacity of $\kappa_{\gamma}=$ 0.03 cm$^2$ g$^{-1}$ \citep{Sutherland:1984}, which is fairly similar to that previously inferred for most magnetar-driven SLSNe \citep{Bersten:2016,Nicholl:2017,Vurm:2021}. Equation 4 of \cite{Bersten:2011} is used to estimate the rate of energy per gram released by radioactive decay, which is the usual expression used in the literature.

To initiate the explosion, a pre-SN model in hydrostatic equilibrium is needed to simulate the conditions of the star prior to the explosion. We adopt the pre-SN models corresponding to single stars with main-sequence masses of 13, 15, 18, 20, and \mbox{25 M$_\odot$} calculated by \cite{Nomoto:1988}. These models are based on a single-star evolutionary code that follows the evolution of the stars from the ZAMS, passing to the He burning, until the core collapse, assuming solar initial abundance. \footnote{No information is available on the host galaxy metallicity (Z) of SN 2023pel. However, GRB-SNe typically originate in low-metallicity environments, suggesting that assuming a lower Z value may be more appropriate. In any case, since our primary focus is to derive the pre-SN properties rather than establish a connection with the ZAMS properties, we do not expect the Z used for the stellar calculation to significantly affect our main results. }. At the moment of the explosion, these stellar models have a He-core of 3.3, 4, 5, 6, and 8 M$_{\odot}$. All these models form a Fe-core before the explosion. The mass coordinate of the Fe nucleus, denoted as M$_{\rm{cut}}$, is assumed to collapse and form a compact remnant during the explosion. Therefore, this mass is removed in our calculation. Specifically, we adopt M$_{\rm{cut}}$ values of approximately 1.4, 1.5, 1.6, 1.7, and 1.8 for the He3.3, He4, He5, He6, and He8 models, respectively, chosen based on the Fe core masses reported by \cite{Tanaka:2009} (see Table 1 of this paper).
Note that although our initial models are not entirely He-free, they will still be used, as we currently lack self-consistent evolutionary models that completely remove the He layers. This approach has also been used in previous studies \citep{Bersten:2016,Taddia:2018}.

Our code simulates the explosion by the injection of a specific amount of energy near the center of the progenitor, for a relatively short duration compared to the hydrodynamic timescale. This energy triggers the formation of a shockwave that propagates through the progenitor and converts thermal and kinetic energy into radiative energy that can be emitted when the shock reaches the stellar surface.  In addition to the usual explosion energy and the energy deposited by the nickel decay, our code also considers the possibility of an extra energy source due to a rapidly rotating and strongly magnetized neutron star (NS), known as a magnetar (see details in \cite{Bersten:2016} and \cite{Orellana:2018jcy}). The rotational energy of a young magnetar, under the assumption of spherical symmetry, is released at a predictable rate, as described by the radiating magnetic dipole \citep{Shapiro:1983}. During the calculation process, when the photosphere recedes deep enough so that part of the magnetar energy is directly deposited outside the photosphere, we add this energy to the bolometric luminosity. This treatment is also used for the nickel deposition \citep{Swartz:1991,Bersten:2016}. We assume that, even if the ejecta is optically thin to optical photons, the magnetar produces hard photons that can be trapped. 

Specifically, the following expressions for the magnetar have been included in the code:

\begin{equation}
    L(t) = \frac{L_p}{(1+\frac{t}{t_p})^2}
\end{equation}
\begin{equation}
    L_p = \frac{4\pi^4R^6}{3c^3} \frac{B^2}{P^4}
\end{equation}
\begin{equation}
    t_p = \frac{3c^3I_{NS}}{2\pi^2R^6} \frac{P^2}{B^2},
\end{equation}

where $I_{NS}= 1 \times 10^{45} \, \rm{g \, cm^2}$ is the moment of inertia of the NS with a radius and mass, assuming a radius of $R=1\times10^6 \, \rm{cm}$ and a mass of $M=1.4$ M$_{\odot}$, respectively. We have also considered an inclination of 30$^\circ$ between the magnetic field and the rotational axis, following previous suggestions (see e.g. \cite{Woosley:2010}).

It has been shown that the brightness of a SN can be significantly increased by the energy deposited from a highly magnetic NS within an expanding SN \citep{Woosley:2010,Kasen:2010}. In fact, this is the most promising source to explain the high luminosity of the SLSNe \citep{Vurm:2021,Hsu:2022}.
Given that SN 2023pel appears to be brighter than normal GRB-SN (see Fig. \ref{fig:model_2023pel} for comparison with SN 1998bw), we will also test the magnetar contribution as an additional energy source in our modeling. The magnetar is mainly characterized by two parameters: the strength of the dipole magnetic field (B) and the initial spin period (P). Alternatively, the spin-down timescale (t$_p$) and the initial rotation energy (E$_{\rm{rot}}$) could also be used \citep{Kasen:2010}. Here, we adopt P and B as the magnetar free parameters to derive in our modeling process.

Our goal is to determine a set of physical parameters that most properly represent the observations of SN 2023pel, by comparing models with observational data. We used the bolometric LC and the Fe II velocities presented in Sect. \ref{sec:Lbol} as observable quantities. As noted earlier, including the velocity information in our modeling is important to reduce the existing degeneracy between the model free parameters \citep{Mazzali:2006, Martinez:2021zbz}. Specifically, the parameters to be determined are: the energy transferred to the envelope after the core collapse (referred to as explosion energy, E), the amount of synthesized radioactive nickel in the explosion (mass of $^{56}$Ni, M$_{\rm{Ni}}$), its degree of mixing (considered as a fraction of the total mass), and the ejected mass related with the pre-SN mass ($\mathrm{M_{preSN}}$) by $\mathrm{M_{ej}}= \mathrm{M_{preSN}} - \mathrm{M_{cut}}$, with $\mathrm{M_{cut}}$  being the mass assumed for the compact remnant formed after the explosion\footnote{In principle, some of the model parameters, such as M$_{\rm{Ni}}$ and its mixing, are provided by the evolutionary calculations. However, because the predicted values often fail to reproduce the data, these parameters are commonly treated as free and are determined by comparison with observations, as considered here.}. In our available grid of models, $\mathrm{M_{preSN}}$ varies between 3.3--8 $\mathrm{M_{\odot}}$,  $\mathrm{M_{cut}}$ in the range of \mbox{1.4--1.85 $\mathrm{M_{\odot}}$}, and the $\mathrm{M_{ej}}$ within 1.9--6 $\mathrm{M_{\odot}}$. Moreover, if the magnetar source is considered, two additional parameters, P and B, must also be included.

\section{Results}\label{sec:results}

\begin{figure*}[h]
    \centering
    \includegraphics[ width=1.0\textwidth]{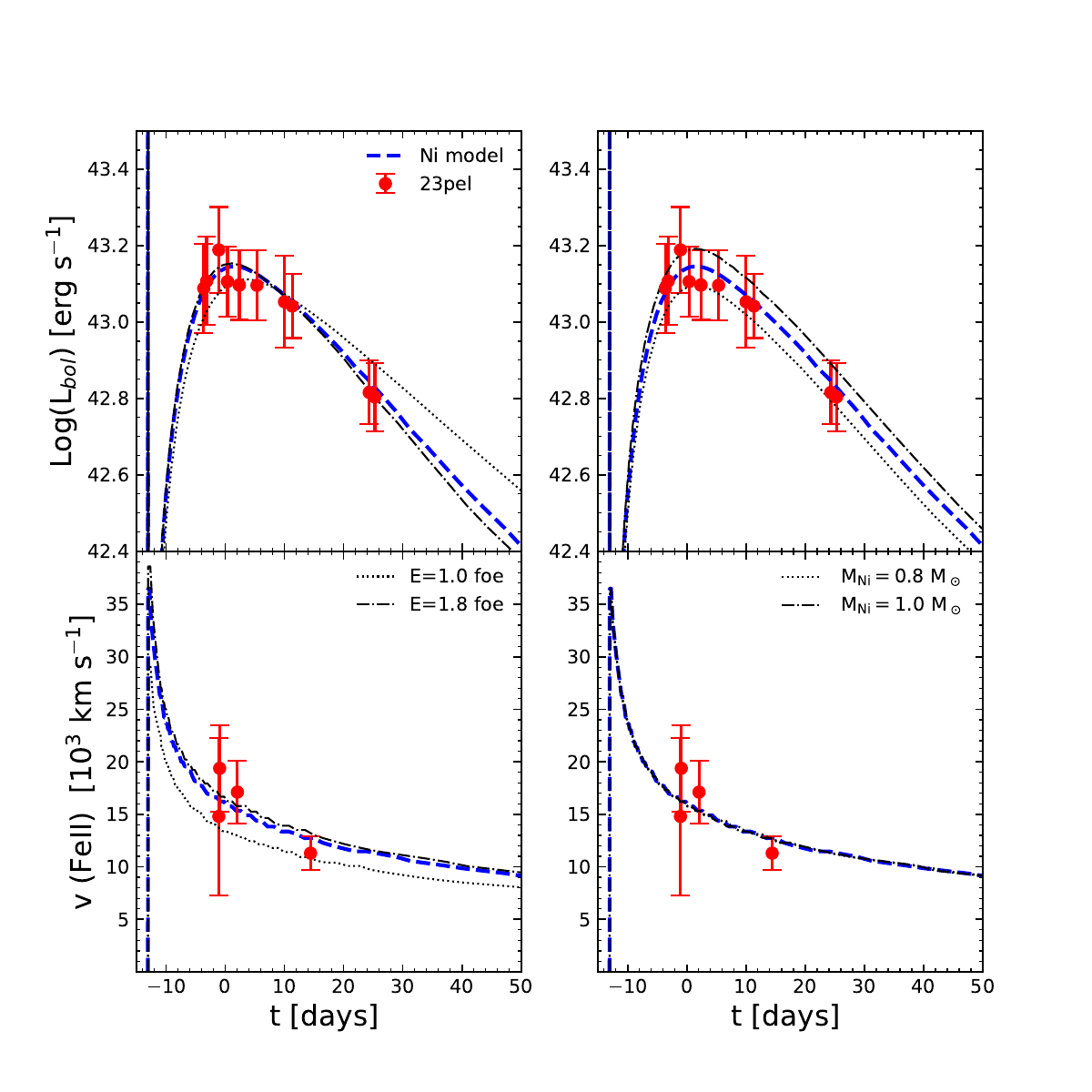}
    \caption{ The best model for the Ni scenario corresponds to the He3.3 model, with an explosion energy of E = 1.6 foe and M$_{\rm{Ni}}$ = 0.9 M$_\odot$ (blue dashed line). Additionally, two more models, with variations on E (left panel) and $\rm{M}_{\rm{Ni}}$ (right panel), are also presented in black lines of different types. These models can also be considered acceptable solutions, thus providing a range of validity for our physical parameters.}
    \label{fig:Ni_model}
\end{figure*}

To search for models that can explain the observations of \mbox{SN 2023pel}, we first consider the five stellar initial configurations presented in Sect. \ref{hidro}. However, only a couple of these pre-SN models were effectively used in a detailed $\chi^2$ inspection, using a wider range of physical parameters to select our preferred option (see more details below).
The reason for this is that our goal is to find a solution that can simultaneously reproduce both the LC and velocity evolution. This requirement restricts our grid of initial models, given the strong dependence of the velocity evolution on explosion energy for a fixed pre-SN mass. It turns out that the energy required to reproduce the velocities in each initial model is not always compatible with the LC rise time and width. These discrepancies cannot always be resolved by adjusting other free parameters, such as the nickel mass and its mixture, constraining the number of initial models to further explore.

We note that the nickel mass has almost no effect on the rise time but strongly affects the $\rm{L}_{\rm{peak}}$. In contrast, the nickel mixing does affect the rise time but does not influence either the $\rm{L}_{\rm{peak}}$ or its width. A precise value of rise time is often uncertain because it depends on accurately estimating the explosion time. However, in the case of SN 2023pel, the explosion time is well-known, as it should be closely coincident with the detection of the GRB. This provides a strong constraint on the rise time and, consequently, on the maximum allowable nickel mixing. In our simulations, we found that full mixing of nickel was always necessary to reproduce the LC data. Therefore, this assumption will be maintained throughout this work.
Everything mentioned above is based on our general knowledge of how the observables (i.e. LC and velocities) respond to variations in the free parameters of the model. These dependencies have been studied and presented in several previous works (e.g. \cite{Bersten:2012,Bersten:2014,Bersten:2023}); therefore, we do not elaborate on them here. As a final point, it is important to mention that the effects on energy described previously also apply in the presence of a magnetar. Hence, also in that case, only a subsample of the initial models has been deeply explored. Nonetheless, since the magnetar provides an additional energy source, the preferred initial model differs from the one of the nickel-powered case. In fact, we found that a more massive model is needed to match the observations in the magnetar scenario. For details on how the considered initial models affect the LC and velocities, see Appendix A.

 With these considerations in mind, in the following subsections we present the best models found for two different scenarios: 1) powered solely by nickel (Sect. \ref{Ni_models}), referred to hereafter as the \emph{Ni model}, and 2) powered by both nickel and a magnetar (Sect. \ref{mag_models}), which will be referred to simply as the \emph{magnetar model} throughout the rest of this paper. However, we highlight that the magnetar model also needs some nickel to reproduce the luminosities.  We found that after $\sim$ 60 days since explosion time, the nickel contribution becomes dominant (see more details in Appendix B).
 
 In each case, once the initial model is selected, we study the variation of the remaining free parameters (E and M$_{\rm{Ni}}$ for the Ni model, and also P and B for the magnetar model). The range of the free parameters considered are E $\in [1,10]$ foe \footnote{1 foe $ = 1 \times 10^{51}$ erg, unit usually used in SN field.} with a step of 0.2 foe, M$_{\rm{Ni}} \in [0.14,1.1]$ M$_\odot$ with a step of 0.02 M$_\odot$, P $\in$ \mbox{$[0.5,15]$ ms} and B $\in$ \mbox{$[2,60] \times 10^{14}\ \rm{G}$} with steps of 0.25 ms and 2 $\times 10^{14}\ \rm{G}$, respectively. This led to a total of 93 models for the Ni case and 180 for the magnetar case. We then calculate the reduced $\chi^2$ (hereafter referred to as $\chi^2$) value for each simulation, considering the data and errors associated with both luminosities and velocities. The models are ranked based on their $\chi^2$ values, with the model having the smallest $\chi^2$  being selected as the preferred choice.

\subsection{Ni models}\label{Ni_models}
In this scenario, the initial model with the smallest mass in our grid, denoted as He3.3, was found to better represent the observations. Specifically, the model has the following parameters: M$_{\rm{ZAMS}}= 13\, $M$_\odot$, $\rm{E}=1.6^{+0.2}_{-0.1}\,\rm{foe}$, $\rm{M}_{\rm{Ni}}=0.90\, \pm{0.15}\, M_\odot$ and an ejecta mass of M$_{\rm{ej}}=$ 1.9 M$_\odot$, considering a M$_{\rm{cut}}$ of 1.4 M$_\odot$. The $\chi^2$ value for this simulation is 3.12, and the reported parameter uncertainties correspond to 1-sigma confidence intervals from the $\chi^2$ distribution. This model is shown with a blue dashed line in both panels of Fig. \ref{fig:Ni_model}. Models with small variations of E (left panel) and M$_{\rm{Ni}}$ (right panel) are also presented.
These models were specifically chosen to fall within the data error bars, thereby providing a range of validity for our parameters. Then, E and $\rm{M}_{\rm{Ni}}$ in the ranges of [1, 1.8] foe and [0.8, 1] M$_\odot$, respectively, can also be considered acceptable solutions\footnote{However, these deviations should not be interpreted as statistical uncertainties. Instead, they represent the effects of single-parameter variations and do not account for potential degeneracy in the solution.}.

As can be seen in Fig. \ref{fig:Ni_model}, models with higher E values enhance the brightness and narrow the LC, affecting both the rise time and the decline rate. Additionally, higher E values lead to higher expansion velocities. On the other hand, M$_{\rm{Ni}}$ only affects the luminosity evolution, with no impact on the velocities. The peak luminosity is extremely sensitive to the nickel mass, reaching higher values for higher nickel masses. However, there is no significant effect on the rise time or the decline rate. Given that the explosion energy significantly affects the velocities, the velocity measurements were crucial for determining this parameter in our models.

\begin{figure}[h]
    \centering
    \includegraphics[width=0.9\linewidth]{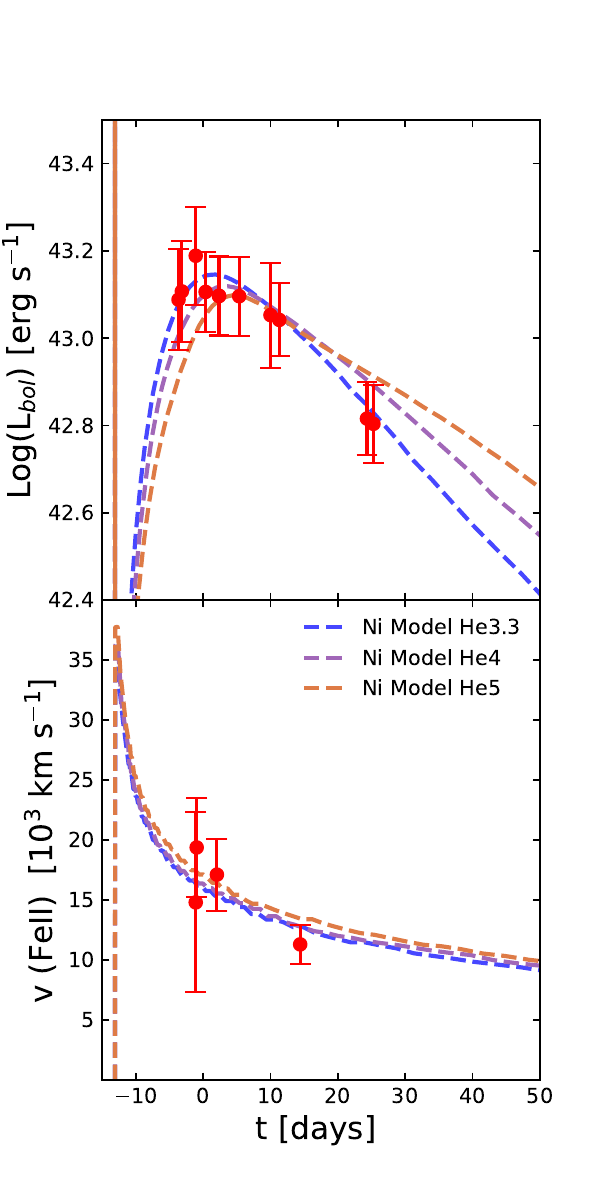}
    \caption{ The optimal model for the Ni scenario (blue dashed line)  compared with two more massive models, He4 (purple dashed line) and He5 (orange dashed line). These models produce less favorable solutions than our preferred model He3.3.}
    \label{fig:best_ni_model}
\end{figure}

In Fig. \ref{fig:best_ni_model} we present a comparison between our optimal Ni model and two more massive models, He4 and He5. The parameters in each case were selected to find the best possible solution for these masses. {Specifically, an explosion energy of \mbox{2 foe} and \mbox{3 foe}, and nickel masses of \mbox{0.93 M$_{\odot}$} and \mbox{0.95 M$_{\odot}$} were found for He4 and He5, respectively.} We note that, although the He4 model could be considered acceptable, the last data points in the LC are marginally within the error bars. This cannot be solved by further increasing the explosion energy due to the low velocities, particularly the velocity at \mbox{$\sim$ 15 days since maximum bolometric luminosity}. A similar difficulty is observed for the He5 model, but the situation is even worse. As a result, the $\chi^2$ values for these models are higher than the He3.3 model, and we could not identify a better solution for more massive models than He3.3 in our grid. However, we cannot rule out models with intermediate masses between He3.3 and He4, whose associated ejected masses would range between 1.9 to 2.5 $\rm{M_{\odot}}$. Therefore, we can consider these values of ejected mass as an acceptable range of validity for our solution.

Finally, we conclude that if SN 2023pel was powered solely by radioactive nickel, its physical parameters should have been approximately an ejected mass of \mbox{1.9 M$_\odot$}, an explosion energy of \mbox{1.6 foe}, and a nickel mass around \mbox{0.90 M$_{\odot}$}. Note that all the acceptable solutions considered, result in high nickel masses, with M$_{\rm{Ni}} \geq$ 0.8 M$_\odot$. Also, the values of the ejecta mass and energy obtained are low compared to other GRB-SNe \citep{Cano:2013,Cano:2017,Kann:2019}.

\subsection{Magnetar models}\label{mag_models}
\label{magnetar}

\begin{figure*}[h]
    \centering
    \includegraphics[ width=1.0\textwidth]{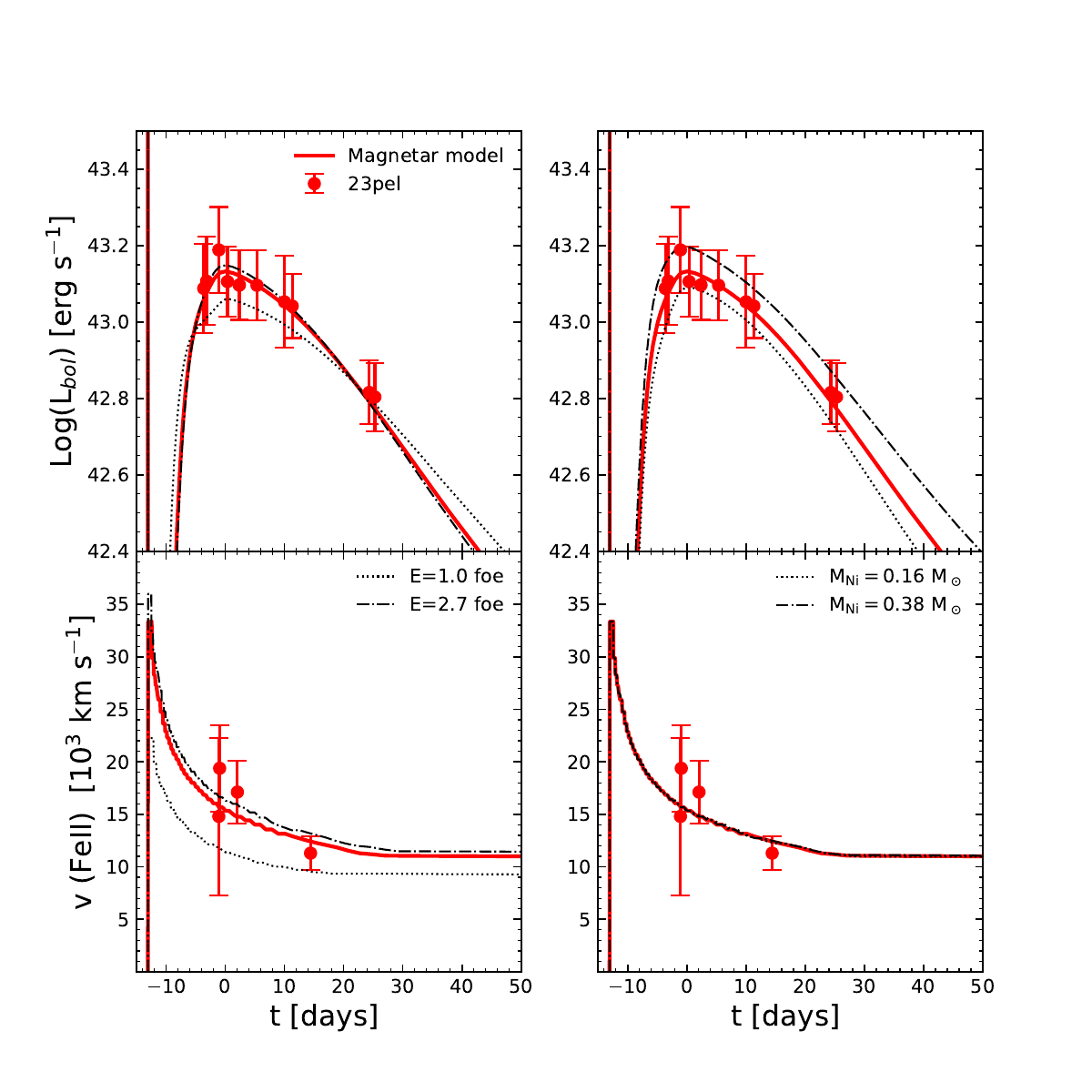}
    \caption{The preferred model for the magnetar scenario, associated with the parameters P$\,=3.2\,$ms,  \mbox{B$\,=28 \times 10^{14}$\,G}, E = 2.3 foe, and M$_{\rm{Ni}}$ = 0.24 M$_{\odot}$, is shown in red solid line. Additionally, two more models with single-parameter variations on  E (left panel) and $\rm{M}_{\rm{Ni}}$ (right panel), are also presented. These models can also be considered acceptable solutions, thus providing a range of validity for our physical parameters.}
    \label{fig:mag_model_1}
\end{figure*}

\begin{figure*}[h]
    \centering
    \includegraphics[ width=1.0\textwidth]{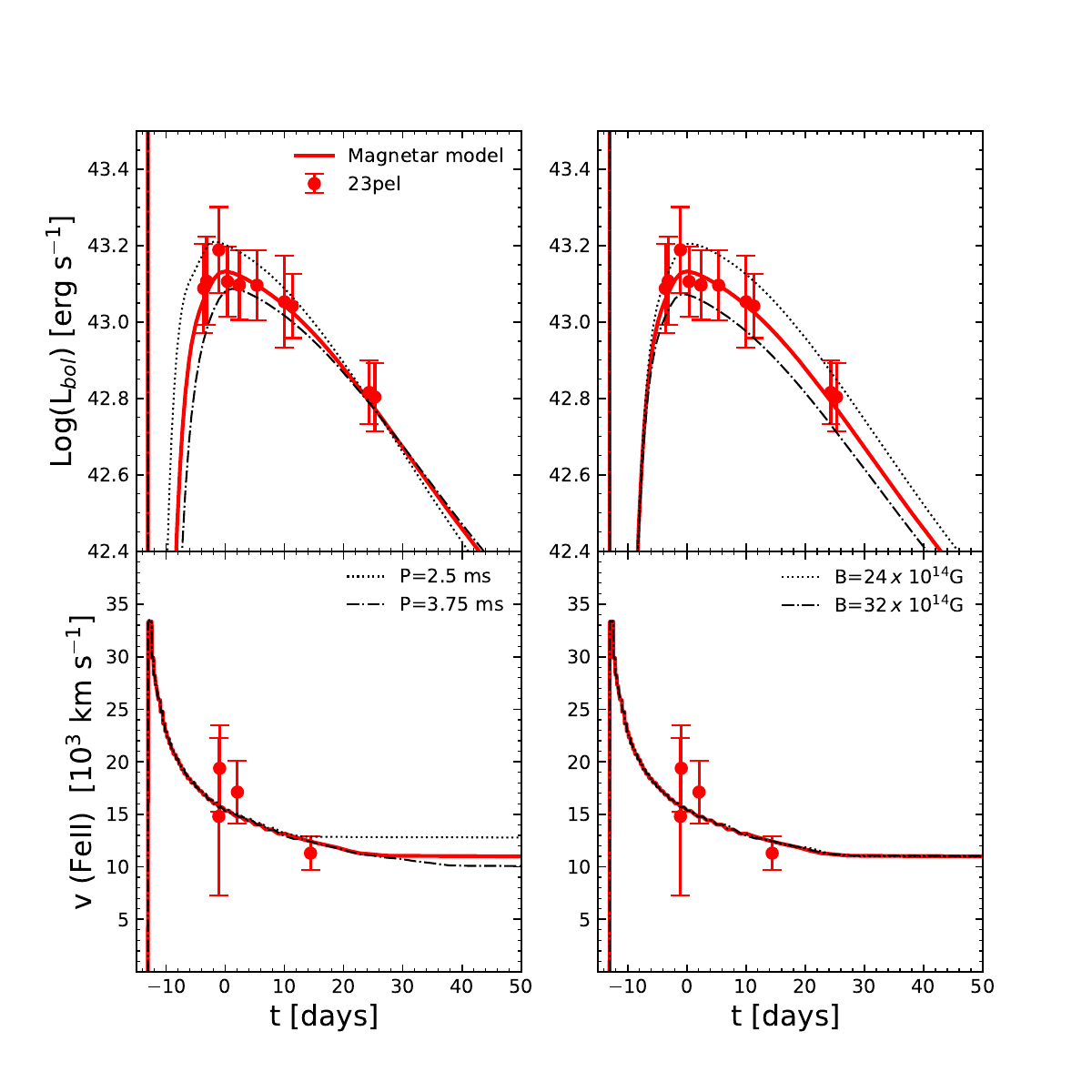}
    \caption{Similar to Fig. \ref{fig:mag_model_1}, but illustrating the effects of varying P (left panel) and B (right panel).}
    \label{fig:mag_model_2}
\end{figure*}

\begin{figure}[h]
    \centering
    \includegraphics[width=0.9\linewidth]{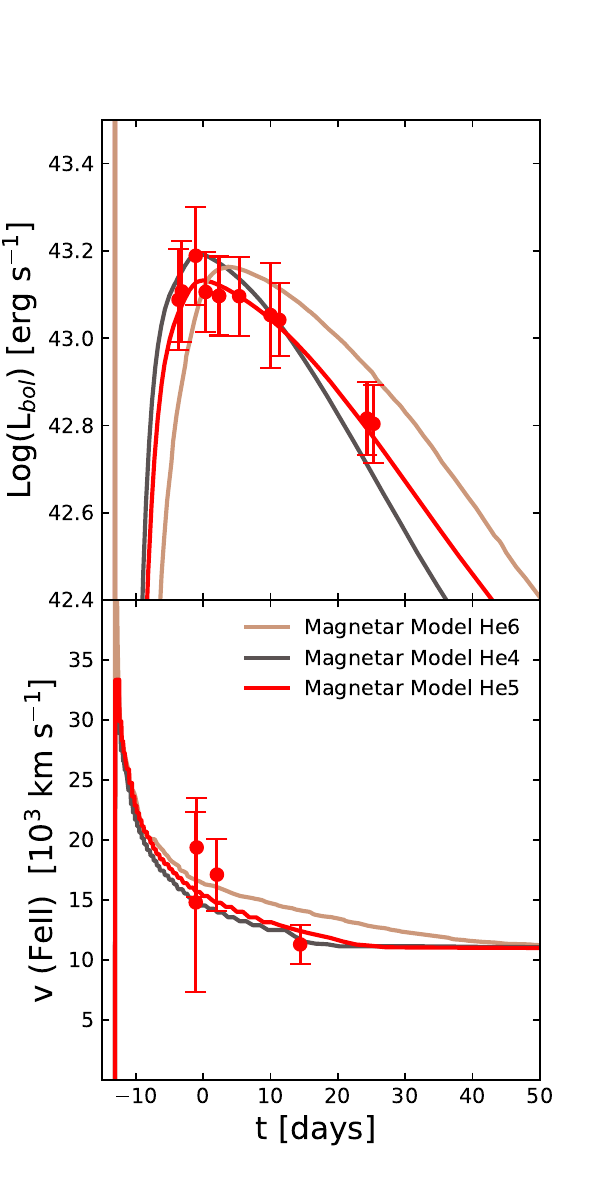}
    \caption{ A comparison between our preferred magnetar model (red solid line) and two more models with pre-SN masses of \mbox{4 $\rm{M}_{\odot}$} (black solid line) and \mbox{6 $\rm{M}_{\odot}$} (gold solid line). It is evident that while the He4 model may be considered acceptable, this is not the case for He6, whose LC does not match the observational data.} 
    \label{fig:best_magnetar_model}
\end{figure}

In this scenario, a more massive pre-SN model (He5) was found to better reproduce the observations of \mbox{SN 2023pel}.
Specifically, the parameters of our preferred model are: \mbox{M$_{\rm{ZAMS}}= 18$ M$_\odot$}, \mbox{E = 2.3$^{+0.3}_{-0.5}\,$foe}, \mbox{M$_{\rm{ej}}= 3.4$ M$_\odot$} considering \mbox{M$_{\rm{cut}}= 1.6$ M$_\odot$}, \mbox{$\rm{P} = 3.2^{+0.1}_{-0.45}$ ms}, and \mbox{$\rm{B}=28^{+0.6}_{-2.0} \times {10^{14}}$\,G}. Additionally, a not negligible amount of nickel with a M$_{\rm{Ni}}= 0.24^{+0.6}_{-0.2}$ $\rm{M}_\odot$ was needed to be included 
to reproduce the luminosity at later times (see details in Appendix B). The uncertainties correspond to 1-sigma confidence intervals derived from the $\chi^2$  analysis. This model has a $\chi^2$ value of 2.6 and is represented by the solid red line in Figs. \ref{fig:mag_model_1}, \ref{fig:mag_model_2} and \ref{fig:best_magnetar_model}.

Figure \ref{fig:mag_model_1} shows our optimal magnetar model (solid red line) alongside other solutions obtained by varying the energy (left panel) and the nickel mass (right panel), while keeping the other parameters fixed. 
The effects of E and M$_{\rm{Ni}}$ on the models are similar to the trend discussed in Sect. \ref{Ni_models} for the Ni model case. However, the changes produced by the E are more moderate, since the presence of the magnetar also influences the energetics and steepens the decline rate.
Considering that the models presented fall within the error bars, they can be used to estimate the validity range of the selected parameters, as discussed in Sect. \ref{Ni_models}. Therefore, models with E and M$_{\rm{Ni}}$ within the ranges $[1,2.7]$ foe and $[0.16,0.38]$ M$_\odot$, respectively, can be considered reasonable solutions.

Similarly, in Fig. \ref{fig:mag_model_2} the effects of varying P (left panel) and B (right panel) on the observables are shown. The spin period affects the peak luminosity, decline rate, and rise time. Lower P values result in a brighter and narrower LC that reaches its peak earlier. P also produces some effect on the expansion velocities, although mainly at later times (t $\gtrsim$ 20 days since maximum bolometric luminosity) where, unfortunately, no observations are available for this object. This latter effect is because the velocity increase is more pronounced in the inner layers, where the extra energy of the magnetar is injected. As the photosphere recedes, it eventually reaches these inner regions, making the dynamic effect more noticeable. Additionally, smaller P values stabilize the velocities at higher levels and earlier times. The envelope is accelerated to high velocities due to the continuous energy injection from the magnetar. Over time, the expansion stabilizes, and most of the material moves at a constant velocity. This final velocity is higher for shorter P values, given the strong dependence of the rotational energy of a NS on the spin period (E$_{\rm{rot}} \propto \rm{P}^{-2}$). On the other hand, variations of B modify the luminosities, while the velocities remain unaffected. Weaker magnetic fields lead to brighter LCs, but they do not influence the rise time or the decline rate. Based on the presented models, we suggest that values of P and B  within $[2.5,3.75]$ ms and $[24,32]\times {10^{14}}$\,G, respectively, can also produce acceptable solutions considering the observational errors. Thus, these values could be considered as validity ranges for the models. 

Note that similar LCs can be obtained for different sets of physical parameters E, P, B, and M$_{\rm{Ni}}$. Again, this underscores the importance of considering the evolution of photospheric velocities, which primarily depend on E and to a lesser extent on P. As a result, using the velocity information was essential for selecting our set of physical parameters. Additionally, P, B and M$_{\rm{Ni}}$ affect the $\rm{L}_{\rm{peak}}$. Therefore, having data around peak luminosity, as we have for SN 2023pel, allows us to improve the determination of these parameters.

Finally, we also explored the possibility of reproducing the observations with other pre-SN masses. In Fig. \ref{fig:best_magnetar_model} we present a comparison between our optimal magnetar model and two additional models, He4 and He6.
Again, all the physical parameters were selected with the intention of ensuring the models fall within the observational error bars as much as possible. Particularly, the explosion energies were found to be \mbox{1.4 foe} and \mbox{3.5 foe}, the nickel masses \mbox{0.1 M$_{\odot}$} and \mbox{0.15 M$_{\odot}$}, the spin periods \mbox{3.7 ms} and \mbox{3.5 ms}, and the magnetic fields \mbox{28 $\times 10^{14}$ G} and \mbox{21 $\times 10^{14}$ G} for He4 and He6, respectively. We note that, while the He4 model could be considered acceptable, this is not the case for the He6 model. The He4 model produces an LC that is marginally outside the error bars around the peak and the radioactive tail. This cannot be solved by increasing the Ni mass, as although this would produce a brighter LC in the tail, as necessary, it would also overestimate the peak luminosity. Alternatively, decreasing the energy would broaden the LC, but it would reduce the velocities, which are already close to the inferior limit of the error bar, thereby producing a worse model. Consequently, the $\chi^2$ value for the He4 model is higher than our preferred He5 model. In the case of He6, it can be seen that the LC is already quite broad and that the peak luminosity does not align with the observations. Although increasing E would correct this problem, it would also result in an overestimation of the velocities. Therefore, the He5 model seems to provide the best solution. Nevertheless, models with pre-SN masses between 4 and 5 M$_\odot$, or ejected masses between 2.5 and 3.4 M$_\odot$ could also be considered acceptable for a reasonable solution.

In conclusion, all the acceptable solutions explored are associated with models involving a millisecond magnetar with a non-negligible nickel contribution. Particularly, our preferred magnetar model for SN 2023pel requires a higher pre-SN mass (He5) compared to the Ni model (He4). This magnetar model suggests that the ejected mass is approximately 3.5 
M$_\odot$, the explosion energy $\sim$ 2.3 
foe, the nickel mass $\sim$ 0.24 
M$_{\odot}$, the spin period $\sim$ 3.2 
ms, and the magnetic field $\sim$ 28 $\times\, 10^{14}$ G. Additionally, our statistical analysis supports the magnetar model over the Ni model.

\section{Discussion and analysis}\label{disc}

In this section, we discuss the viability of the models for SN~2023pel found in the previous section, and compare our results with other models available in the literature. Then, we make a comparison with other objects to study the overall physical parameters for this GRB-SN, providing insights into its possible progenitor. Finally, we briefly discuss the physical caveats of our analysis.

\subsection{Analysis of the preferred models}

\begin{figure}[h]
    \centering
    \includegraphics[ width=0.5\textwidth]{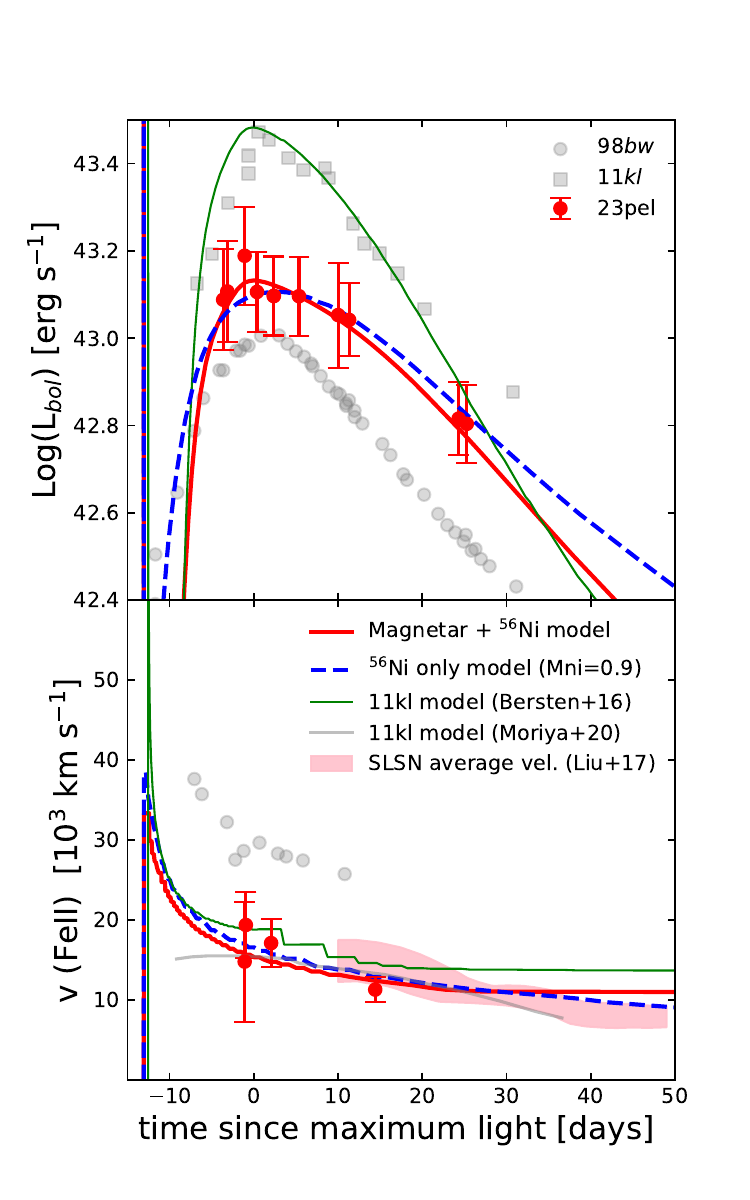}
    \caption{Comparison between observations and models. Top panel: Bolometric LCs. Bottom panel: Photospheric and line
     velocities evolution. Time is shown in days since maximum bolometric luminosity. The red solid line represents our preferred model for SN 2023pel, associated with the lowest $\chi^2$, driven by a magnetar with P$\, =$ 3.2 ms, {B$\, =28 \times 10^{14}$ G and $\mathrm{M_{Ni}}= 0.24  \,\mathrm{M_{\odot}}$ (see Sect. \ref{sec:results} for more parameter details).} The blue dashed line represents the Ni model, with $\mathrm{M_{Ni}}= 0.9  \,\mathrm{M_{\odot}}$. Solid green and gray lines represent models for SN 2011kl calculated by \cite{Bersten:2016} and \cite{Moriya:2020cxt}, respectively. The shaded area denotes the weighted average velocities of SLSNe Ic (pink) calculated by \cite{Liu:2017}. 
     Observations of SN 1998bw \citep{Clocchiatti:2011} and SN 2011kl \citep{Greiner:2015} are included for comparison.} 
    \label{fig:model_2023pel}
\end{figure}

In Fig. \ref{fig:model_2023pel}, the two preferred models described in Sect. \ref{sec:results} are shown together. For comparison, the famous SN 1998bw, and the luminous SN 2011kl are also included. In addition, models for SN 2011kl published in \cite{Bersten:2016} and \cite{Moriya:2020cxt} are also presented. SN 2011kl is the brightest GRB-SN recorded to date, although it is still less luminous than SLSNe which are associated with $\rm{L}_{\rm{bol}}>3\times 10^{44}$ erg s$^{-1}$ \citep{Gal-Yam:2019}. From Fig. \ref{fig:model_2023pel}, it is evident that \mbox{SN 2023pel} is less luminous than \mbox{SN 2011kl}, but brighter than \mbox{SN 1998bw}. The studies conducted by S24 and HD24 reported values of L$_{\rm{peak}} \sim 1.3 \times 10^{43}\, \rm{erg}\,\rm{s}^{-1}$ and L$_{\rm{peak}} \sim 5.75 \times 10^{42}\, \rm{erg}\,\rm{s}^{-1}$, respectively. 
Although the first case shows good agreement with our results, their value was obtained using a different bolometric calculation method. In the second case, the discrepancy is more significant, and we believe it is closely related to the fact that HD24 computes pseudo-bolometric luminosities using a limited wavelength range, without applying flux corrections. 

A comparison between the expansion velocities of different SNe is presented in the lower panel of Fig. \ref{fig:model_2023pel}. The shaded pink area included in the same figure represents the weighted average velocities of the hydrogen-poor SLSNe calculated by \cite{Liu:2017}. The same study also provides average velocities for SNe Ic-BL; however, these are not included in Fig. \ref{fig:model_2023pel}, as the sample does not distinguish between events with and without an associated GRB. Given that GRB-SNe exhibit systematically higher velocities \citep{Modjaz:2016}, we aimed to avoid any potential confusion. In all cases, the time is shown in days since the maximum bolometric luminosity. We note that \mbox{SN 2023pel} has velocities that are significantly lower than those of the \mbox{SN 1998bw} and are closer to the expected values for SLSNe.

Although we found two satisfactory solutions, we consider the magnetar model to be the most probable. While it has the smallest $\chi^{2}$ value, the difference $\Delta \chi^{2}$ = 3.12 - 2.6 = 0.52 is not statistically significant, suggesting that both models provide similarly good fits to the data. However, Ni models always require high nickel masses ($\sim 0.9$ M$_{\odot}$) to reproduce the high luminosity observed for \mbox{SN 2023pel}.
Such large amounts of nickel production are difficult to explain, particularly given the lower ejecta masses (M$_{\rm{ej}}\sim$ 2 M$_\odot$) found in our analysis. Moreover, there is a known tension between nucleosynthesis models and nickel mass estimates for GRB-SNe. \citet{Suwa:2019} reported that the expected Ni masses for core-collapse SNe range from 0.07 to 0.28 M$_\odot$, with the higher values corresponding to more massive progenitors, as expected for GRB-SNe. Similarly, \citet{Leung:2023} found that for very massive progenitors, the predicted Ni masses range from 0.1 to 0.3 M$_\odot$. These findings suggest that nucleosynthesis alone cannot account for Ni masses of 0.5 M$_\odot$ or higher. For our preferred Ni model, a ratio of M$_{\rm{Ni}}/\rm{M}_{\rm{ej}} = 0.47$  is obtained. This value is significantly higher than those typically inferred for most GRB-SNe, which usually have values around 0.07 \citep{Cano:2013}). Therefore, our preference for the magnetar model is primarily based on these physical considerations. A similar difficulty was found in the analysis of SN~2011kl, where an additional energy source was proposed to explain its high luminosity \citep{Greiner:2015,Bersten:2016, Kann:2016ndk}. Furthermore, this aligns with prior research suggesting that highly luminous GRB-SNe, luminous or superluminous SNe may be linked to additional energy sources, such as magnetars \citep{Woosley:2010,Dessart:2012,Wang:2016,Soker:2017,Afsariardchi:2020glh,Gomez:2022,Kumar:2024, Gottlieb:2024}. In general, the aforementioned studies have determined distinct ranges of magnetar parameters for SLSNe and GRB-SNe. Specifically, SLSNe usually require lower values of spin period and magnetic field, with P $\in$ [1,10] ms and B $\in$ [0.1,10] $\times 10^{14}$ G \citep{Woosley:2010,Wang:2016,Soker:2017,Nicholl:2017,Chen:2023,Gottlieb:2024,Gomez:2024}. In contrast, those necessary for GRB-SNe are higher, with P $\in$ [5,35] ms and B $\in$ [10,100] $\times 10^{14}$ G \citep{Dessart:2012,Wang:2016,Afsariardchi:2020glh,Song:2023qdv,Kumar:2024}. Our simulations are consistent with these trends since higher values of P and B result in less luminous LCs (see Fig. \ref{fig:mag_model_2}). However, the parameters that we derived for SN 2023pel ($\rm{P}=3.2$ ms and $\rm{B}=28 \times 10^{14}$ G; see Sect. \ref{mag_models}), fall within an intermediate range between these two groups. This highlights the peculiarity of this object and suggests that SN 2023pel may represent a transitional event between typical GRB-SNe and SLSNe.

Additionally, the exploration and election of the magnetar model were also motivated by findings in prior works about \mbox{SN 2023pel}. As previously mentioned, S24 modeled the LC of \mbox{SN 2023pel} using the semi-analytic code Arnett \citep{Arnett:1982}. The low values obtained for some of their parameters (\mbox{M$_{\rm{Ni}}=$ 0.38 $\pm$ 0.01 M$_\odot$}, \mbox{$\rm{E}\sim 1.3\,$foe}, and M$_{\rm{ej}}=$ 1.0 $\pm$ 0.6 M$_\odot$) caught our attention and were one of the motivations for this study. Note that our Ni model required a significantly larger value of nickel mass \mbox{($\rm{M}_{\rm{Ni}} \sim 0.9$)}. We believe that the primary cause of this discrepancy is the difference in bolometric luminosity calculations. Furthermore, differences in LC modeling and the absence of photospheric velocity evolution in the analysis may also be a contributing factors. However, the differences in the ejected mass and explosion energy are comparatively smaller, given the weaker dependence of these parameters on absolute luminosity.  

S24 also tested a millisecond magnetar scenario for \mbox{SN 2023pel}, following the model outlined by \cite{Cano:2016iuo}.
The magnetar model employed by S24 estimates a total luminosity comprising three contributions: (a) the central engine, (b) the GRB afterglow, and (c) the SN component (multiplied by a normalization factor $\Phi$). However, when attempting to fit the X-ray and optical observations separately, the parameters characterizing the magnetar do not match with each other. The normalization factor is employed to scale the model to the optical data. They argue that even if the parameters match, $\Phi > 1$ implies the necessity of an additional power source to account for the observed flux. They found $\Phi \sim 8$ and concluded that a model solely powered by a magnetar is not feasible, suggesting the requirement for a supplementary power source, as we also found here (see details in S24).

In this work, we searched for a magnetar model focused exclusively on the SN component, employing a one-dimensional hydrodynamic code. In our case, the magnetar parameters (P and B) and the other free parameters (E, mix, M$_{\rm{Ni}}$) were derived simultaneously, attempting to reproduce both the bolometric LC and the expansion velocities. Our approach does not require a normalization factor, as our code inherently accounts for different contributions to luminosity, including energy from the central magnetar and synthesized radioactive nickel. Notably, the preferred magnetar model we identified includes a non-negligible contribution from radioactive nickel (see Sect. \ref{mag_models}).

Additional information would be valuable for distinguishing between nickel- and magnetar-powered models. For instance, prompt emission or early X-ray plateaus have been associated with energy contributions from a millisecond magnetar \citep{Cano:2014,Metzger:2015}. Emission from the pulsar wind nebula in radio, X-rays, and gamma rays may also indicate the presence of a magnetar \citep{Kotera:2013,Metzger:2014,Omand:2018,Murase:2021,Omand:2024}, as may higher ionization lines in the spectrum \citep{Chevalier:1992,Omand:2023,Dessart:2024}. \cite{Poidevin:2023} proposed that an increase in polarization during post-maximum phases may indicate photospheric asymmetries driven by magnetar dynamics. Also, late-time photometric observations (t > 60 days) could further assist in differentiating the models, as their LC slopes diverge noticeably.  Furthermore, spectra dominated by nickel are expected to display distinct features compared to those powered by a magnetar.
Although such observations would provide crucial insights into the dominant power source and help refine our understanding of the underlying mechanisms, they are not currently accessible for SN 2023pel.

\subsection{Comparison with other objects}\label{sec:comparison}

\begin{figure}[h]
    \centering
    \includegraphics[ width=0.5\textwidth]{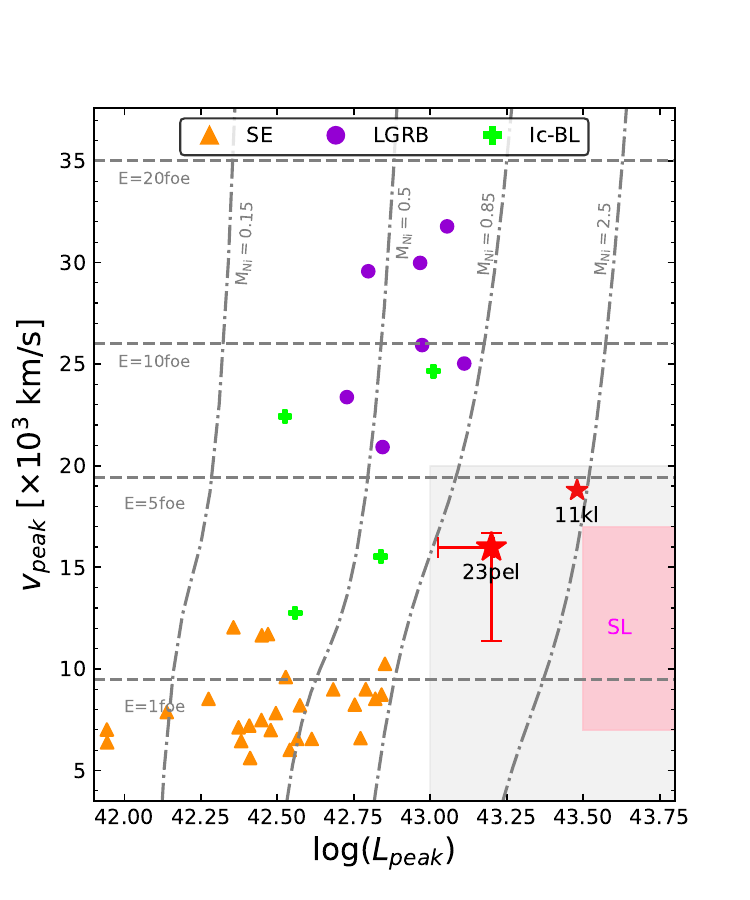}
    \caption{ L$_{\rm{peak}}$ vs. v$_{\rm{peak}}$ plane. SN~2023pel and SN~2011kl are marked by red stars. A sample of 7 GRB-SNe \citep[purple dots]{Saez:2024}, 31 SE SNe \citep[orange triangles]{Taddia:2018}, and 4 SNe classified as Ic-BL but not associated with GRBs \citep[green crosses]{RomanAguilar:23,Taddia:2018} are included for comparison. The pink region indicates the expected location of SLSNe. The grey area indicates a region of high luminosities (L$_{\rm{peak}}$) and low v$_{\rm{peak}}$, where Ni models require M$_{\rm{Ni}} \gtrsim 0.85 \rm{M}_\odot$.
     We also show Ni models with fixed energy but varying M$_{\rm{Ni}}$ (dashed lines), and fixed M$_{\rm{Ni}}$ but varying E (dot-dashed lines), for an ejecta mass of 3.4 M$_\odot$.
    }
    \label{fig:grillas}
\end{figure}

In Fig. \ref{fig:grillas}, we present a comparison of the photospheric velocities at the time of maximum bolometric luminosity (v$_{\rm{peak}}$) versus the L$_{\rm{peak}}$ for different SNe. For each object, t$_{\rm{peak}}$, L$_{\rm{peak}}$, and v$_{\rm{peak}}$ were measured from the corresponding optimal model. In addition to SN 2023pel and SN 2011kl, we have included for comparison a sample of 7 GRB-SN \footnote{The 7 GRB-SNe correspond to: SN 1998bw/GRB 980425, SN 2003dh/GRB 030329, SN 2003lw/GRB 031203, SN 2006aj/XRF 060218, SN 2012bz/GRB 120422, SN 2013dx/GRB 130702A, and SN 2016jca/GRB 161219B. The data were extracted from \cite{Cano:2017, Taddia:2019, Ashal:2019, Cano:2017B, DElia:2015, Schulze:2014}.} recently analyzed by \mbox{Saez et al. (in prep.)}, 31 SE SNe from \cite{Taddia:2018}, and 4 SNe Ic-BL of which three were studied in \cite{RomanAguilar:23} (SN 2010qts, SN 2014dby and SN 2020bvc) and one in \cite{Taddia:2018} (SN 2009bb). The selection of these objects was based on the following criteria: (i) all have good temporal, photometric, and spectroscopic coverage, ensuring reliable modeling, and (ii) they were modeled using the same hydrodynamic code used here (see Sect. \ref{hidro}), facilitating direct comparisons. Consequently, the sample of each SN subtype used for comparison was limited to those that meet our criteria. Furthermore, we have shaded in pink the region of L$_{\rm{peak}}$ and v$_{\rm{peak}}$ where SLSNe are expected to be located \citep{Liu:2017,DeCia:2017uqv}. In Fig. \ref{fig:grillas}, we also present Ni models with varying E values while keeping the M$_{\rm{Ni}}$ fixed (dot-dashed lines), and Ni models with different M$_{\rm{Ni}}$ values keeping the E fixed (dashed lines). These models help to identify the E and M$_{\rm{Ni}}$ values required to reach the v$_{\rm{peak}}$ and L$_{\rm{peak}}$ values of SN 2011kl and SN 2023pel.

From Fig. \ref{fig:grillas}, it can be seen that SN 2023pel and \mbox{SN 2011kl} stand out from the rest, positioning themselves in a region of high $\rm{L}_{\rm{peak}}$ ($\gtrsim 10^{43}\,$erg s$^{-1}$) and relatively low v$_{\rm{peak}}$ \mbox{($\lesssim 20 \times 10^3 \, \rm{km}\, \rm{s}^{-1}$)}. Previously, SN 2011kl has been considered an outlier compared to other SLSNe, being less luminous, evolving faster, and residing within the "luminosity gap" between GRB-SNe and SLSNe \citep{Kann:2019}. We note that \mbox{SN 2023pel} exhibits similar characteristics, and both objects may represent a transition between the rapidly expanding GRB-SNe and the extremely luminous SLSNe. This analysis suggests that the \mbox{L$_{\rm{peak}}$-v$_{\rm{peak}}$} plane may be used to distinguish between objects with potentially different power sources (see Fig. \ref{fig:grillas}). Moreover, the same figure indicates that for the Ni models, reaching the grey region where \mbox{SN 2023pel} and \mbox{SN 2011kl} are located requires high nickel masses ($\gtrsim 0.85$ M$_\odot$) and low energies \mbox{($\lesssim$ 5 foe)}. This suggests that modeling objects in this region may require an additional energy source beyond radioactive nickel. 

\subsection{Physical caveats}

We have found viable solutions to reproduce the observations of SN 2023pel. The magnetar model seems to be the most plausible explanation for the high luminosities and low velocities, as we discussed in Sect. 5.1. In particular, unlike semi-analytic models (\cite{Nicholl:2017,Omand:2024}), the range of masses and chemical compositions are constrained by stellar evolutionary calculations. Nevertheless, our hydrodynamic models offer significant advantages, such as the use of realistic opacity tables, an appropriate equation of state, progenitors derived from stellar evolution models, and a more detailed treatment of the photospheric velocity, which allow for a more physically consistent comparison of both luminosity and velocity. 
Besides, our code employs a simplified treatment of radiative transfer, a diffusion approximation for optical photons, and a gray approximation for gamma rays, allowing reliable comparisons only with bolometric luminosities rather than color LCs (see \cite{Wang:2022} for details on the limitations of fitting bolometric LCs instead of multiband LCs). In addition, our 1-D approach does not account for possible asymmetries that might be expected in this type of explosion. Specifically, 2-D simulations indicate lower effective opacities due to hydrodynamic instabilities, which result in brighter but shorter LC peaks (\citealp[Fig. 9]{Suzuki:2021}). Consequently, a 1-D simplification may lead to an underestimation of the total ejected mass. We have also assumed that the magnetar energy is fully deposited and thermalized in the ejecta (that is, the magnetar does not lose energy and that all of its energy is deposited where it is injected). These are among the most relevant limitations of our code.

Another important aspect to consider is the choice of the initial models used to initialize the explosion. As mentioned before, these are He-rich structures calculated from stellar evolution models that assume solar metallicity. While deviations in metallicity are expected to be minor and primarily affect the connection between pre-SN and ZAMS properties, the unavailability of He-free models is a more significant issue. However, to the best of our knowledge, there is no stellar calculation that can self-consistently remove the entire He layer. Given the absence of realistic He-free stellar models, we have decided to employ our well-studied pre-SN model grid rather than introduce hand-crafted modifications to the models, without a solid physical justification. Despite these limitations, it is important to highlight that this is the first numerical study of the SN 2023pel available in the literature. Additional numerical models of SN 2023pel would be very useful for verification and comparison with our results.

\section{Conclusions}\label{sec:conclus}

We have computed the bolometric LC of SN~2023pel, after applying the corresponding K-corrections. Our results indicate that this SN is very luminous, reaching L$_{\rm{peak}}\sim 1.5 \times 10^{43}\, \rm{erg \, s^{-1}}$. Subsequently, we have performed the hydrodynamical modeling of the bolometric LC and photospheric velocity evolution of SN~2023pel, enabling us to derive comprehensive physical parameters. Our findings reveal that, compared to other detected GRB-SNe, SN~2023pel is highly luminous and exhibits lower expansion velocities. These intrinsic characteristics suggest a scenario where a millisecond magnetar is the primary power source, complemented by a non-negligible contribution of radioactive nickel. When considering a model powered solely by Ni, the required Ni masses to reproduce the observed high luminosities (M$_{\rm{Ni}}>0.80 \, \rm{M}_{\odot}$) exceed those typically expected for GRB-SNe. Consequently, our preferred model involves a magnetar with a spin period of 3.2$^{+0.1}_{-0.45}$ ms, a magnetic field of \mbox{28$^{+0.6}_{-2.0}$$\, \times 10^{14}$ G}, explosion energy of approximately 2.3 $^{+0.3}_{-0.5}$ foe, a nickel mass of 0.24$^{+0.6}_{-0.2}$ M$_{\rm{\odot}}$, and an ejected mass around 3.5 M$_{\rm{\odot}}$. The uncertainties correspond to 1-sigma confidence intervals derived from the $\chi²$ analysis, and their respective ranges of validity are discussed in Sects. \ref{Ni_models} and \ref{mag_models}. 

Finally, we have compared the L$_{\rm{peak}}$ and v$_{\rm{peak}}$ values of SN~2023pel with those of various groups of SE SNe. From this comparison, we have found that SN~2023pel occupies a distinct region from other GRB-SNe, aligning more closely with those classified as SLSNe. The \mbox{L$_{\rm{peak}}$-v$_{\rm{peak}}$} plane could be helpful to distinguish between objects in regions of high luminosities and low velocities, which would seem to indicate the necessity of extra energy sources to avoid relying on unrealistically large masses of nickel. We have also noted that SN~2023pel presents a similar behavior to SN~2011kl, associated with an ultra-long duration GRB, and being the most luminous GRB-SN observed thus far. Interestingly, SN~2011kl was also explained by considering a magnetar to account for its high luminosity. 

Note that, although we have explored two different models, there are other possible mechanisms that could explain the observables of SN 2023pel. One such mechanism is the circumstellar matter (CSM) interaction (see e.g. \cite{Suzuki:2018,Gomez:2021,Leung:2021,Salmaso:2023,Konyves-Toth:2025}); however, no clear evidence of CSM interaction has been found in its spectrum. The possible existence of fallback material onto the compact remnant \citep{Salmaso:2023,Niblett:2025} or even the possibility that the jet itself deposits additional energy \citep{Soker:2024,Kumar:2025} could influence the dynamics. Although these scenarios are worth considering, they have not been explored in this study.

It would be valuable to verify our results with a larger sample of GRB-SNe and to study the characteristics that may help to decide the most suitable formation scenario for these objects. Including more of these SNe could be crucial in determining whether this observed difference is genuine or merely a statistical anomaly.

\begin{acknowledgements}
   We extend our gratitude to the National Research Council of Argentina (CONICET). L.M.R.A. and K.E. are Doctoral Fellows of CONICET, while M.B. is a member of the Scientific Research Career of CONICET. M.M.S. is awaiting her admission to CONICET as an assistant researcher, a position awarded in 2022. We also thank Gokul Srinivasaragavan for providing the photometric data of SN~2023pel. We would like to acknowledge the Supernova Observations and Simulations (S.O.S.) group from FCAG-UNLP, particularly Dr. Gaston Folatelli, for the valuable and enriching discussions that greatly facilitated the development of this research. M.M.S. expresses her gratitude to iTHEMS at RIKEN, the NSF Network for Neutrinos, Nuclear Astrophysics, and Symmetries (N3AS) Physics Frontier Center at U.C. Berkeley, and the RIKEN Astrophysical Big Bang Laboratory for providing a supportive research environment. Additionally, M.M.S. was supported by the NSF under cooperative agreement 2020275 and by the Japan Science and Technology Agency (JST) as part of the Adopting Sustainable Partnerships for Innovative Research Ecosystem (ASPIRE), Grant Number JPMJAP2318. 
\end{acknowledgements}

%
   \bibliographystyle{aa} 
   \bibliography{biblio} 
%

\begin{appendix}

\onecolumn
\section{Variation of the initial model}\label{appA}

\begin{figure}[h]
    \centering
    \includegraphics[width=0.4\textwidth]{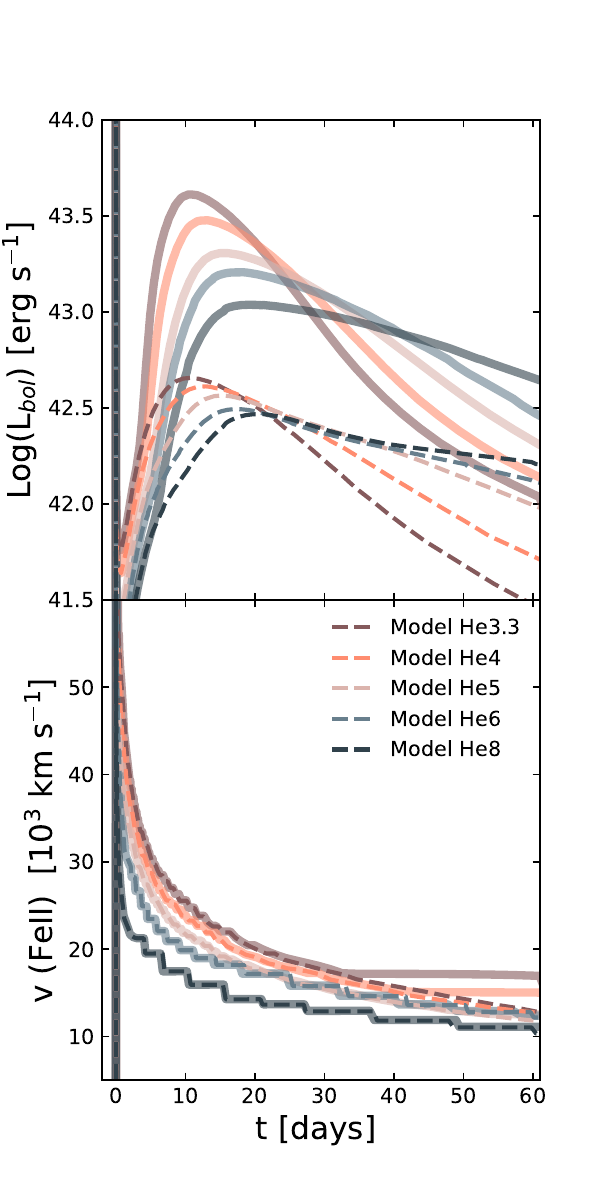}
    \caption{Light curves (upper panel) and velocities evolution (lower panel) as a function of the pre-SN model. Ni models with $\rm{E}=5\, \rm{foe}$, $\rm{M}_{\rm{Ni}}= 0.25 \, \rm{M}_{\odot}$ are shown with dashed lines. Magnetar models with $\rm{E}=5\, \rm{foe}$, $\rm{M}_{\rm{Ni}}= 0.25 \, \rm{M}_{\odot}$, $\rm{P}=4\, \rm{ms}$ and $\rm{B}=21 \times 10^{14}\, \rm{G}$ are shown with solid lines.}
    \label{fig:models_variation}
\end{figure}

This appendix presents the effects of varying pre-SN masses on the LCs and photospheric velocities. Our available grid of initial models includes masses of 3.3, 4, 5, 6, and 8 M$_{\rm{\odot}}$, as detailed in Sect. \ref{hidro}. For each simulation we used the same explosion and magnetar parameters, to focus only on the effect of the pre-SN mass. Specifically, we used: \mbox{E$\,=\,$5 foe}, \mbox{M$_{\rm{Ni}}=$ 0.25 M$_{\rm{\odot}}$}, \mbox{P$\,=\,$4 ms}, \mbox{B$\,=\,$ 21 $\times 10^{14}$ G} and a fully Ni mixing in all cases. The results are presented in Fig. \ref{fig:models_variation}. Ni models are represented by dashed lines, while solid lines indicate models that also include the contribution of the magnetar. From the figure, we can highlight several
characteristics: 1) More massive models produce less luminous and wider LCs, leading to longer rise times and shallower decline rates. Conversely, higher pre-SN masses result in lower photospheric velocities over time. These patterns are observed in all models, regardless of whether they include a magnetar. 2) A slightly different behavior is observed in the evolution of velocities: from a certain time onward, the decline stabilizes in the presence of the magnetar. This stabilization occurs at later times for more massive models.
3) Although the magnetar critically affects the luminosity, its effect on velocity is much subtler.

All the trends identified in this appendix were considered when selecting the most appropriate initial model for each scenario studied in this work.

\section{Ni contribution to the magnetar model.}

\begin{figure}[h]
    \centering
    \includegraphics[width=0.4\textwidth]{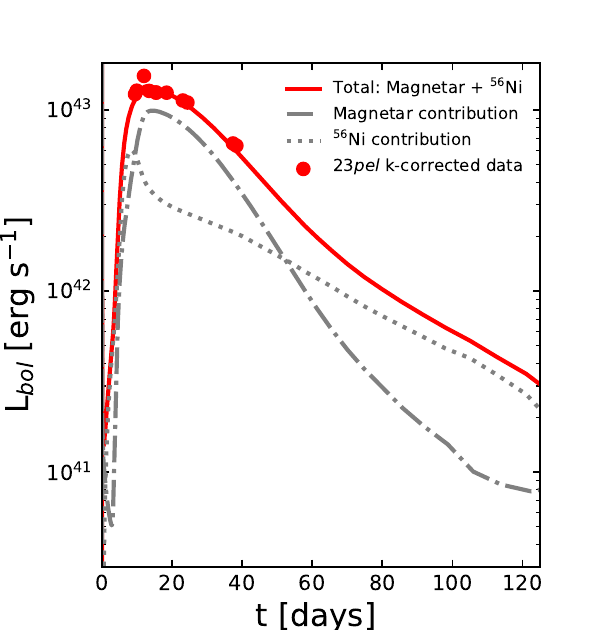}
    \caption{LC of the preferred model (red solid line) of SN 2023pel. 
    The dot-dashed line corresponds to an LC of the magnetar model without considering radioactive material, while the other parameters have the same values as those in the solid line. The dotted line represents the difference between these two LCs, indicating the contribution of $^{56}$Ni.}
    \label{fig:model_descompuesto}
\end{figure}

In Fig. \ref{fig:model_descompuesto}, we show the preferred magnetar model found for SN~2023pel and the different contributions to the luminosity. Here, we aim to present the effects on the LC resulting from the inclusion of Ni as a complementary energy source to the magnetar.
 The LC produced exclusively by the magnetar contribution is shown in a dot-dashed line, while the LC generated only by $^{56}$Ni is shown in a dotted line. As seen in Fig. \ref{fig:model_descompuesto}, the Ni contribution predominantly affects the luminosity at later times. Meanwhile, before $\sim$ 60 days, the magnetar provides the main contribution to the bolometric luminosity. The plot shows the essential role of $^{56}$Ni in powering the late-time bolometric LC. It is important to note that late-time observations for this type of objects could be helpful in more precisely determining the necessary amount of Ni and corroborating the models.

\end{appendix}
\end{document}